\documentclass[a4paper, twoside, reqno, 12pt, dvips]{amsart}
\RequirePackage{fix-cm} 

\usepackage{fixltx2e}     

\usepackage{amssymb}
\usepackage{amsmath}
\usepackage{mathrsfs, dsfont}

\usepackage{a4wide}

\usepackage{acronym}

\usepackage{indentfirst}
\usepackage{graphicx}
\usepackage{psfrag}

\usepackage[usenames,dvipsnames]{pstricks}
\usepackage{epsfig}
\usepackage{pst-grad} 
\usepackage{pst-plot} 

\usepackage{subfigure}

\usepackage{longtable}

\usepackage[french,english]{babel}
\usepackage[latin1]{inputenc}

\usepackage{color}
\definecolor{oneblue}{rgb}{0,0.0,0.75}
\usepackage[colorlinks,
            urlcolor=oneblue,
            linkcolor=oneblue,
            citecolor=oneblue,
            bookmarksopen=false,
            pagebackref]{hyperref}

\numberwithin{equation}{section}

\newcommand{\phis}{\ensuremath{\varphi}}
\newcommand{\R}{\ensuremath{\mathbb{R}}}
\newcommand{\M}{\ensuremath{\mathcal{M}}}
\newcommand{\N}{\ensuremath{\mathcal{N}}}
\newcommand{\A}{\ensuremath{\mathcal{A}}}
\newcommand{\K}{\ensuremath{\mathbf{K}}}
\newcommand{\ud}{\ensuremath{\mathrm{d}}}
\newcommand{\F}{\ensuremath{\mathcal{F}}}
\renewcommand{\S}{\ensuremath{\mathcal{S}}}
\newcommand{\eps}{\ensuremath{\varepsilon}}
\renewcommand{\H}{\ensuremath{\mathcal{H}}}
\renewcommand{\L}{\ensuremath{\mathcal{L}}}
\newcommand{\Hilb}{\ensuremath{\mathbf{H}}}

\newcommand{\dx}{\ensuremath{\partial_x}}
\newcommand{\dX}{\ensuremath{\partial_X}}
\newcommand{\od}[2]{\ensuremath{\frac{\ud #1}{\ud #2}}}
\newcommand{\pd}[2]{\ensuremath{\frac{\partial #1}{\partial #2}}}

\begin{document}

\title[Solutions to a compact deep-water equation]{Special solutions to a compact equation for deep-water gravity waves}

\author[F. Fedele]{Francesco Fedele$^*$}
\address{School of Civil and Environmental Engineering \& School of Electrical and Computer Engineering, Georgia Institute of Technology, Atlanta, USA}
\email{fedele@gatech.edu}
\urladdr{http://savannah.gatech.edu/people/ffedele/Research/}
\thanks{$^*$ Corresponding author}

\author[D. Dutykh]{Denys Dutykh}
\address{LAMA, UMR 5127 CNRS, Universit\'e de Savoie, Campus Scientifique, 73376 Le Bourget-du-Lac Cedex, France}
\email{Denys.Dutykh@univ-savoie.fr}
\urladdr{http://www.lama.univ-savoie.fr/~dutykh/}

\begin{abstract}
Recently, Dyachenko \& Zakharov (2011) \cite{Dyachenko2011} have derived a compact form of the well known Zakharov integro-differential equation for the third order Hamiltonian dynamics of a potential flow of an incompressible, infinitely deep fluid with a free surface. Special traveling wave solutions of this compact equation are numerically constructed using the Petviashvili method. Their stability properties are also investigated. In particular, unstable traveling waves with wedge-type singularities, viz. peakons, are numerically discovered. To gain insights into the properties of these singular solutions, we also consider the academic case of a perturbed version of the compact equation, for which analytical peakons with exponential shape are derived. Finally, by means of an accurate Fourier-type spectral scheme it is found that smooth solitary waves appear to collide elastically, suggesting the integrability of the Zakharov equation.
\end{abstract}


\maketitle

\tableofcontents

\section{Introduction}

The Euler equations that describe the irrotational flow of an ideal incompressible fluid of infinite depth with a free surface are Hamiltonian. Their symplectic formulation was discovered by \textsc{Zakharov} (1968) \cite{Zakharov1968} in terms of the free-surface elevation $\eta(x,t)$ and the velocity potential $\phis(x,t) = \phi(x, z = \eta(x,t),t)$ evaluated at the free surface of the fluid. Here, $\eta(x, t)$ and $\phis(x, t)$ are conjugated canonical variables with respect to the Hamiltonian $\H$ given by the total wave energy. It is well known that the Euler quations are completely integrable in several important limiting cases. For example, in a two-dimensional (2-D) ideal fluid, unidirectional weakly nonlinear narrowband wave trains are governed by the Nonlinear Schr\"odinger (NLS) equation, which is integrable \cite{Zakharov1972}. Integrability also holds for certain equations that models long waves in shallow waters, in particular the Korteweg--de Vries (KdV) equation (see, for example, \cite{Ablowitz1974, Ablowitz1981, John, Whitham1999}) or the Camassa--Holm (CH) equation \cite{Camassa1993}. For these equations, the associated Lax-pairs have been discovered and the Inverse Scattering Transform unveiled the dynamics of solitons, which elastically interact under the invariance of an infinite number of time-conserving quantities \cite{Ablowitz1974, Ablowitz1981, John, Whitham1999}.

Another important limiting case of the Euler equations for the free-surface of an ideal flow is that considered by \cite{Zakharov1968, Zakharov1999}. By means of a third order expansion of $\H$ in the wave steepness, he derived an integro-differential equation in terms of canonical conjugate Fourier amplitudes, which has no restrictions on the spectral bandwidth. To derive the Zakharov (Z) equation, fast non-resonant interactions are eliminated via a canonical transformation that preserves the Hamiltonian structure \cite{Krasitskii1994, Zakharov1999, Gramstad2011}. The integrability of the Z equation is still an open question, but the fully nonlinear Euler equations are non-integrable \cite{Dyachenko1994, Dyachenko1996b}. Indeed, non-integrability can be easily proven by considering the terms of the perturbation series of the Hamiltonian in powers of the wave steepness limited on their resonant manifolds. To have non integrability, it is enough to prove that at least one of these amplitudes is nonzero. In this regard, \cite{Dyachenko1994, Dyachenko1996b} conjectured that the Z equation for unidirectional water waves (2-D) is integrable since the nonlinear fourth-order term of the Hamiltonian vanishes on the resonant manifold, leaving only trivial wave-wave interactions, which just cause nonlinear frequency shifts of the separate Fourier modes. They also pushed their analysis to the next order and proved that the effective fifth-order term does not vanish on the corresponding resonant manifold. Thus, the Euler equations of the free-surface hydrodynamics are not integrable in two dimensions \cite{Dyachenko1996b}.

Recently, \textsc{Dyachenko} \& \textsc{Zakharov} (2011) \cite{Dyachenko2011} realized that the trivial resonant quartet-interactions that occur in the 2-D free surface dynamics could be further removed by a canonical transformation. As a result, the Z equation drastically simplifies to the compact form
\begin{equation}\label{eq:cDZ}
  ib_t = \Omega b + \frac{i}{8}\Bigl(b^\ast\bigl((b_x)^2\bigr)_x - \bigl(b_x^\ast(b^2)_x\bigr)_x\Bigr) - \frac{1}{4}\Bigl[b\K\{|b_x|^2\} - \bigl(b_x\K\{|b|^2\}\bigr)_x\Bigr],
\end{equation}
where $b$ is a canonical variable, $b_t$, $b_x$ denote partial derivatives with respect to $t$ and $x$	and the symbols of the pseudo-differential operators $\Omega$ and $\K = \Hilb[\partial_x]$ ($\Hilb$ being the Hilbert transform) are given, respectively, by $\sqrt{g|k|}$ and $|k|$, $k$ being the Fourier transform parameter. Further, $b$ relates to the wave surface $\eta$ as 
\begin{equation}\label{eq:1.2}
  \eta = \sqrt{\frac{\omega_0}{2g}}a + c.c.,
\end{equation}
where $a$ is a conjugate canonical variable that relates to $b$ via a non-trivial canonical transformation (see \cite{Dyachenko2011}), and \emph{c.c.} denotes complex conjugation. To leading order in the wave steepness $a = b$.

In this study, we explore (\ref{eq:cDZ}), hereafter referred to as cDZ, for analytical studies and numerical investigations of special solutions in the form of solitary waves. The present paper is organized as follows. First, we derive the envelope equation associated to cDZ and then smooth ground states and traveling waves are numerically computed using the Petviashvili method (\cite{Petviashvili1976}, see also \cite{Lakoba2007}). Further, unstable traveling waves with wedge-type singularities, viz. peakons, are also numerically identified. To gain insights into the properties of singular traveling waves we consider a perturbed version of the local compact equation, where the non-local terms involving the operator $\K$ are neglected in (\ref{eq:cDZ}). For this academic case we are able to derive analytical peakons with exponential shape. Finally, the interaction of smooth traveling waves is numerically investigated by means of an accurate Fourier-type pseudo-spectral scheme.

\section{Envelope dynamics and special solutions}

Consider the following ansatz for wave trains in deep water
\begin{equation}\label{eq:ansatz}
  b(X,T) = \eps\sqrt{\frac{2g}{\omega_0}} a_0 B(X,T) e^{i(k_0x - \omega_0 t)},
\end{equation}
where $B$ is the envelope of the carrier wave $e^{i(k_0x - \omega_0 t)}$, and $X=\eps k_0(x-c_g t)$, $T = \eps^2\omega_0 t$, with $k_0 = \frac{\omega_0^2}{g}$ and $\omega_0$ as characteristic wavenumber and frequencies. The small parameter $\eps = k_0a_0$ is a characteristic wave steepness associated to an amplitude $a_0$ and $c_g$ is the wave group velocity in deep water. From (\ref{eq:1.2}), the leading order wave surface $\eta$ is given in terms of the envelope $B$ as
\begin{equation}\label{eq:etaEnv}
  \eta(X,T) = \eps a_0 B(X,T) e^{i(k_0x - \omega_0 t)} + \mathrm{c.c.}
\end{equation}
Using (\ref{eq:ansatz}) in (\ref{eq:cDZ}) yields the cDZ envelope equation
\begin{eqnarray}\label{eq:cDZenv}
  iB_T &=& \Omega_\eps B + \frac{i}{4}\bigl(B^\ast\S((\S B)^2) + iB^\ast(\S B)^2 - 2\S\bigl(B|\S B|^2\bigr)\bigr) \nonumber \\
  && - \frac{\eps}{2}\Bigl[B\K\{|\S B|^2\} - \S\bigl(\S B\K\{|B|^2\}\bigr)\Bigr],
\end{eqnarray}
where $\S = \eps\partial_X + i$. The approximate dispersion operator $\Omega_\eps$ is defined as follows
\[
  \Omega_\eps := \frac18\partial_{XX} + \frac{i}{16}\eps\partial_{XXX} - \frac{5}{128}\eps^2\partial_{XXXX} + \frac{7i}{256}\eps^3\partial_{XXXXX},
\]
where $o(\eps^3)$ dispersion terms are neglected. Equation (\ref{eq:cDZenv}) admits three invariants, viz. the action $\A$, momentum $\M$ and the Hamiltonian $\H$ given, respectively, by
\begin{equation}\label{eq:H}
  \H = \int_\R\Bigl[B^\ast\Omega_\eps B + \frac{i}{4}|\S B|^2[B(\S B)^\ast - B^\ast\S B] - \frac{\eps}{2}|\S B|^2\K(|B|^2)\Bigr]\,\ud X,
\end{equation}
\begin{equation}\label{eq:AM}
  \A = \int_\R B^\ast B\,\ud X, \qquad 
  \M = \int_\R i\bigl(B^\ast\S B - B(\S B)^\ast\bigr)\,\ud X.
\end{equation}

Expanding the operator $\S$ in terms of $\eps$, (\ref{eq:cDZenv}) can be written in the form of a generalized derivative NLS equation as
\begin{equation}\label{eq:BT}
iB_T = \Omega_\eps B + |B|^2B - 3i\eps|B|^2B_X - \frac{\eps}{2}B\K\{|B|^2\} + \eps^2 \N_2(B) + \eps^3 \N_3(B) = 0,
\end{equation}
where $B_X$ denotes differentiation with respect to $X$, 
\begin{eqnarray*}
  \N_2(B) &=& -\frac{3}{2}B^\ast(B_X)^2 + B|B_X|^2 - |B|^2B_{XX} + \frac12B^2B_{XX}^\ast + \nonumber \\
  && \frac{i}{2}\Bigl[B\K(B^\ast B_X - BB_X^\ast) + B_X\K|B|^2 + \bigl(B\K|B|^2\bigr)_X\Bigr],
\end{eqnarray*}
and
\[
 \N_3(B) = -\frac{i}{2}|B_X|^2B_X + \frac{i}{2}B_{XX}(B^\ast B_X - BB_X^\ast) - \frac{i}{2}B B_X B_{XX}^\ast - \frac12\Bigl[B\K|B_X|^2 - \bigl(B_X\K|B|^2\bigr)_X\Bigr].
\]
To leading order (\ref{eq:BT}) reduces to the NLS equation.

\subsection{A Hamiltonian Dysthe equation}

Hereafter, we will study a special case of (\ref{eq:BT}), which is a symplectic version of the temporal Dysthe equation (\cite{Dysthe1979}, see also \cite{Stiassnie1984, Stiassnie1984a, Hogan1985, Trulsen1997}). Keeping terms to $O(\eps)$ in (\ref{eq:BT}) yields
\begin{equation}\label{eq:cDZDysthe}
   iB_T = \Bigl(\frac{1}{8}\partial_{XX} + \frac{i\eps}{16}\partial_{XXX}\Bigr)B + |B|^2B - 3i\eps|B|^2B_X - \frac12\eps B\K|B|^2,
\end{equation}
hereafter referred to as cDZ-Dysthe. Note that the original temporal Dysthe equation is not Hamiltonian since expressed in terms of multiscale variables, which are usually non canonical (see, for example, \cite{Fedele2011}). This is common in mathematical physics. Typically, the dynamics is governed by partial differential equations expressed in terms of physically-based variables, which are not usually canonical. A transformation to new variables is needed in order to unveil the desired structure explicitly (see, for example, \cite{Seliger1968}). This is the case for the equations of motion for an ideal fluid: in the Eulerian description, they cannot be recast in a canonical form, whereas in a Lagrangian frame the Hamiltonian structure is revealed by Clebsch potentials (see, for example, \cite{Seliger1968, Morrison1998}). Moreover, multiple-scale perturbations of differential equations expressed in terms of non-canonical variables typically lead to approximate equations that do not maintain the fundamental conserved quantities, as the hydrostatic primitive equations on the sphere \cite{Lorenz1960}, where energy and angular momentum conservation are lost under the hydrostatic approximation.

As an example, consider the finite dimensional system of an harmonic oscillator in the classical canonical variables $q(t)$ (coordinate) and $p(t)$ (momentum). This admits the canonical form 
\begin{equation}\label{eq:ho}
  \dot{q} = \pd{\H}{p} = p, \qquad \dot{p} = -\pd{\H}{q} = -q,
\end{equation}
where $\dot{p}$ denotes time derivative, and $\H = (q^{2}+p^{2})/2$. The flow in the phase-space is 'incompressible' since the divergence vanishes:
\[
\pd{\dot{q}}{q} + \pd{\dot{p}}{p} = 0.
\]
The transformation $z=q+ip$ is canonical and (\ref{eq:ho}) transforms to
\begin{equation}\label{eq:h1}
  \dot{z} = -i\pd{\H}{z^{\ast}} = -iz, \qquad 
  \dot{z}^{_{\ast }} = i\pd{\H}{z}=iz^{\ast},
\end{equation}
where $\H = \left\vert z\right\vert^{2}/2$, and $z^{\ast }$ is the complex conjugate of $z$. It is straightforward to prove that the gauge transformation 
\[
  z = w e^{i\alpha\left\vert w\right\vert ^{2}},
\]
with $\alpha$ as a free parameter, is also canonical, and (\ref{eq:h1}) remains unchanged in the new variables $w$ and $w^\ast$. On the other hand, if one considers the coordinate change
\begin{equation}\label{eq:zn}
  Q = \frac{q}{\sqrt{1 + \alpha q^{2}}}, \qquad P = p,
\end{equation}%
then (\ref{eq:ho})\ transforms to the noncanonical form
\begin{equation}\label{eq:w}
  \dot{P} = -Q\left( \sqrt{1 + \alpha P^{2}}-2\alpha P^{2}\right) ,\qquad 
  \dot{Q} = \frac{P}{\sqrt{1 + \alpha P^{2}}}.
\end{equation}
This flow does not preserve volume as (\ref{eq:ho}) does, nonetheless equations (\ref{eq:w}) are those of a disguised harmonic oscillator obtained via the noncanonical change of variables (\ref{eq:zn}) (see also \cite{Morrison1998}). The original Dysthe equation \cite{Dysthe1979} shares the same roots as (\ref{eq:w}). They both come from a noncanonical transformation of a Hamiltonian system. For the spatial version of the Dysthe equation, canonical variables have been identified by means of a gauge transformation, and the hidden Hamiltonian structure is unveiled \cite{Fedele2011}. The more general canonical transformation proposed by \cite{Dyachenko2011} yields the symplectic form (\ref{eq:cDZDysthe}) of the Dysthe equation \cite{Dysthe1979}.

In the following, insights into the underlying dynamics of the cDZ equation and associated Dysthe equation are to be gained if we construct some special families of solutions in the form of ground states and traveling waves of the envelope $B$, often just called solitons or solitary waves, as described below.

\subsection{Ground states and traveling waves}\label{sec:gs}

To begin, consider the cDZ equation (\ref{eq:cDZenv}). We construct numerically ground states and traveling waves (TWs) of the envelope $B$ of the form $B(X, T) = F(X - cT)e^{-i\omega T}$, where the function $F(\cdot)$ is in general complex, $\omega$ and $c$ are dimensionless frequency and velocity of the TW with respect to a reference frame moving with the group velocity $c_g$. In physical space the true frequency and velocity are given, respectively, by $\tilde\omega = \eps^2\omega_0\omega$ and $\tilde{c} = \eps c\omega_0/k_0$. After substituting this ansatz into the governing equation (\ref{eq:cDZenv}) we obtain the following nonlinear steady problem for $F$(in the moving frame $X - cT$)
\[
\L F = \N(F),
\]
where $\L = \omega - ic - \Omega_\eps$ and $\N(F)$ denotes nonlinear terms of the cDZ of (\ref{eq:cDZenv}). In order to solve this equation we use the Petviashvili method \cite{Petviashvili1976, Pelinovsky2004, Lakoba2007, Yang2010}. This numerical approach has been successfully applied in deriving TWs of the spatial version of the Dysthe equation \cite{Fedele2011}. Schematically, the iteration takes the following form
\[
  F_{n+1} = \S^\gamma \L^{-1}\cdot \N(F_n), \quad
  \S = \frac{\langle F_n, \L\cdot F_n\rangle}{\langle F_n, \N(F_n)\rangle},
\]
where $\S$ is the so-called stabilyzing factor and the exponent $\gamma$ is usually defined as a function of the degree of nonlinearity $p$ ($p = 3$ for the cDZ equation). The rule of thumb prescribes the following formula $\gamma = \frac{p}{p-1}$. The scalar product is defined in the $L_2$ space. The inverse operator $\L^{-1}$ can be efficiently computed in the Fourier space. To initialize the iterative process, one can use the analytical solution to the leading order NLS equation (see the associated Dysthe equation (\ref{eq:cDZDysthe})). We point out that this method can be very efficiently implemented using the Fast Fourier Transform (FFT) (see, for example, \cite{Frigo2005}).

\begin{figure}
  \centering
  \subfigure%
  {\includegraphics[width=0.49\textwidth]{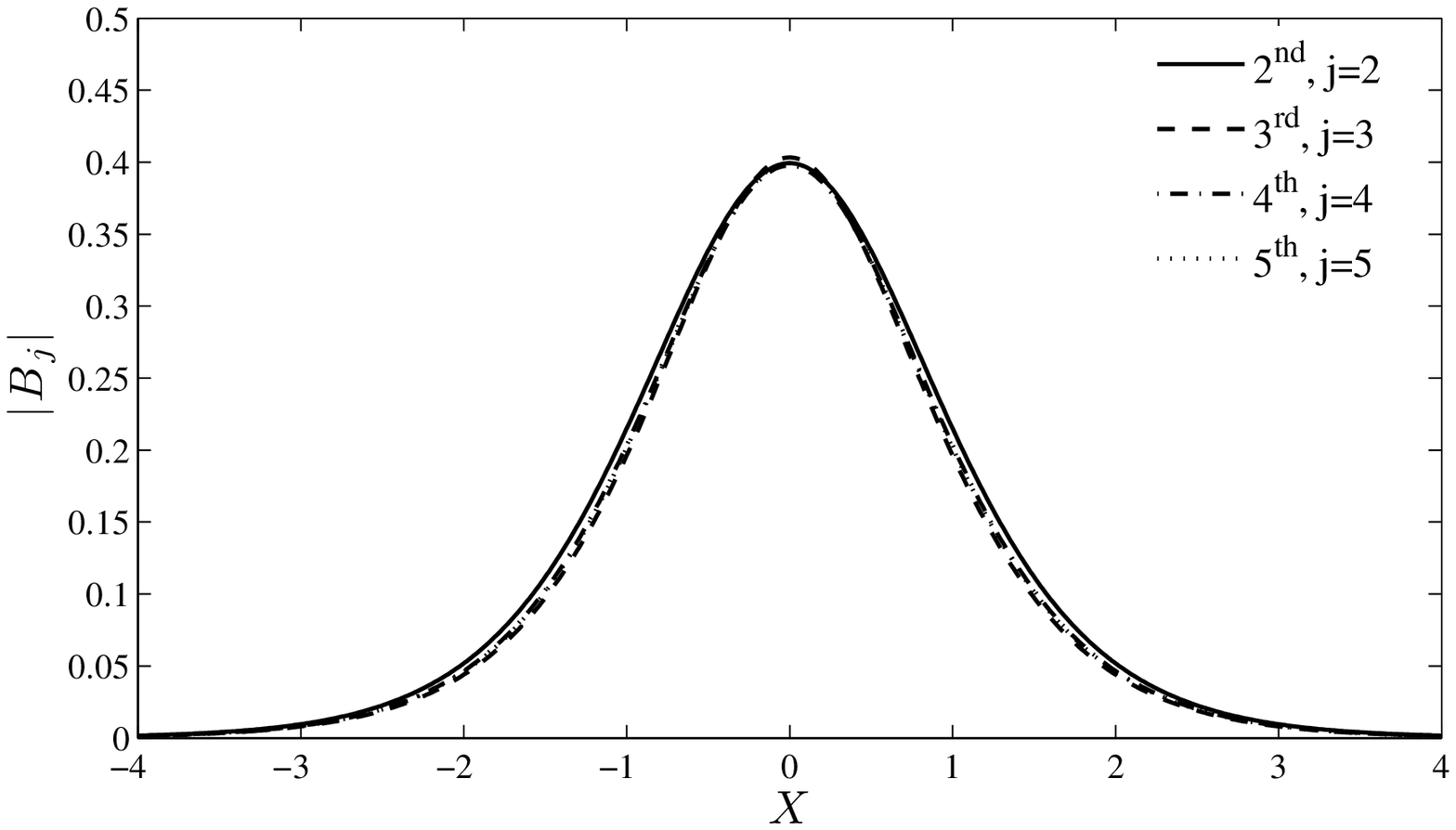}}
  \subfigure%
  {\includegraphics[width=0.49\textwidth]{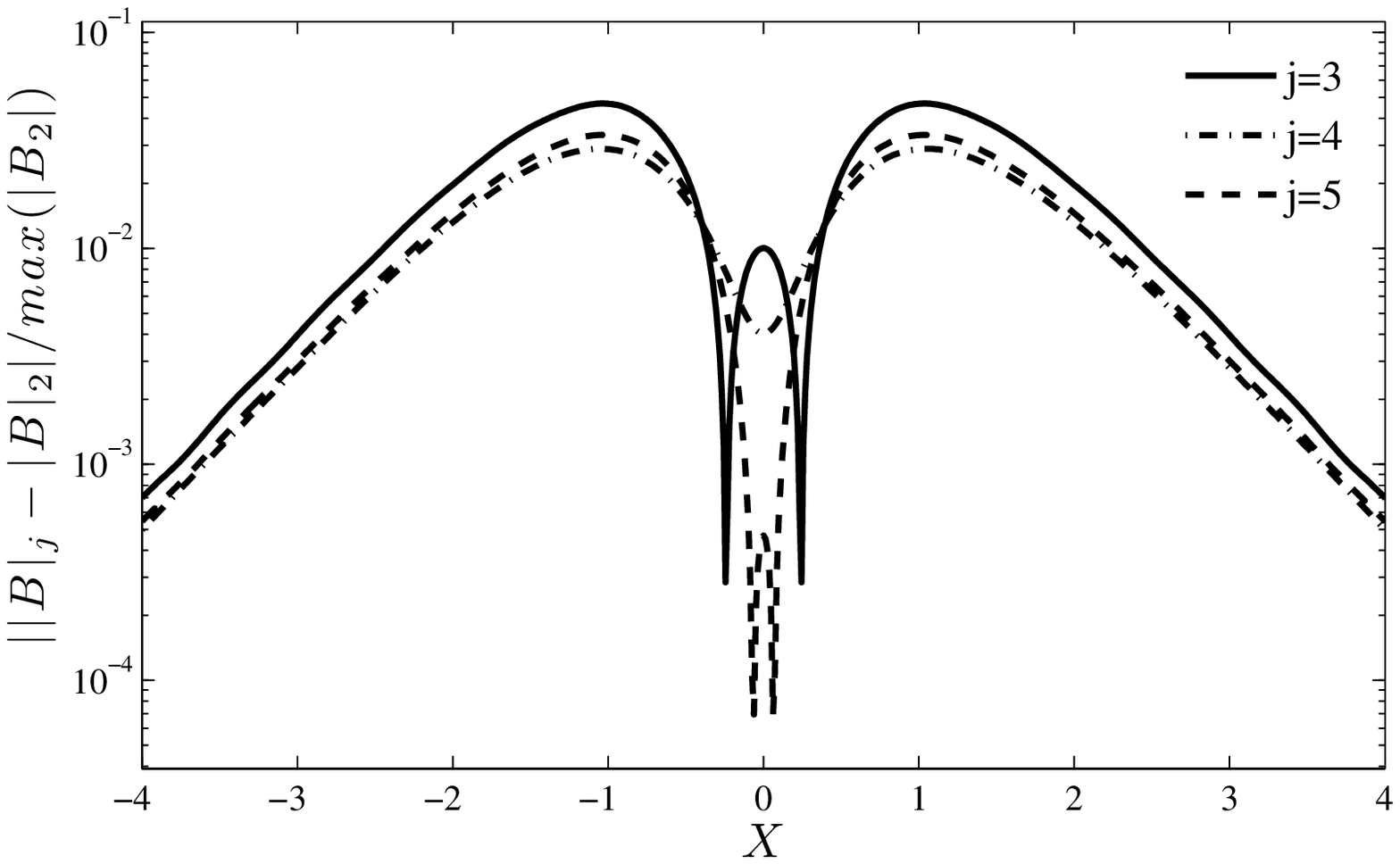}}
  \caption{cDZ equation: Convergence of the ground state $B_j$ ($\eps = 0.19$, $\omega = 0.1$, $c = 0$) with respect to the approximate dispersion relation that retains terms up to second, third, fourth and fifth orders ($j = 2, \ldots, 5$). Left: soliton shapes; right: normalized differences $\bigl||B_j| - |B_2|\bigr|$ for $j = 3,4,5$.}
  \label{fig:fig0}
\end{figure}

\begin{figure}
  \centering
  \includegraphics[width=0.99\textwidth]{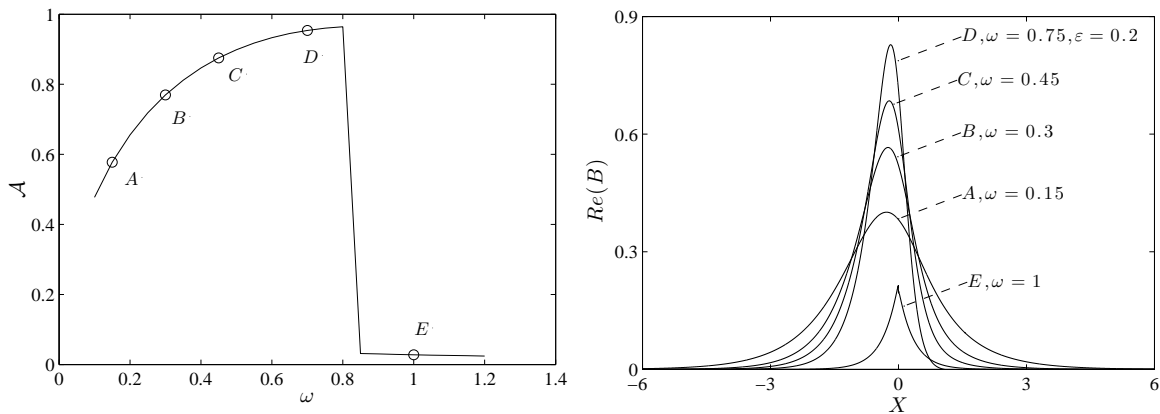}
  \caption{cDZ equation: (left) Action $\A$ of ground states of the envelope $B$ as function of the frequency $\omega$ and (right) bifurcation of a peakon from smooth ground states ($\eps = 0.2$, $c = 0$).}
  \label{fig:fig1}
\end{figure}

\begin{figure}
  \centering
  \includegraphics[width=0.99\textwidth]{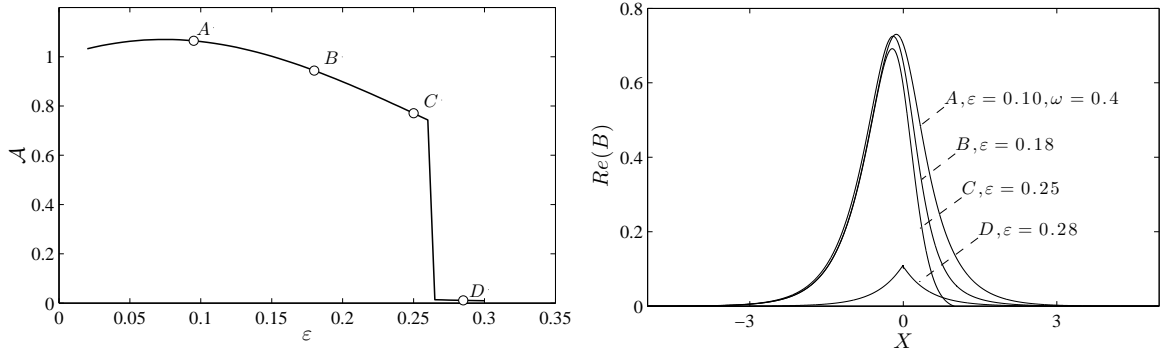}
  \caption{cDZ equation: (left) Action $\A$ of ground states of the envelope $B$ as function of the steepness $\eps$ and (right) bifurcation of a peakon from smooth ground states ($\omega = 0.4$, $c = 0$).}
  \label{fig:fig1_5}
\end{figure}

Without loosing generality, hereafter we just consider the leading term of the dispersion operator, viz. $\Omega_\eps = \frac18\partial_{XX}$, since the soliton shape is only marginally sensitive to the higher order dispersion terms as shown in Figure \ref{fig:fig0}. We also observed that the Petviashvili method does not converge if we retain the full dispersion operator. Figure \ref{fig:fig1} shows the action $\A$ of the ground states ($c = 0$, but moving with the group velocity in the physical frame) as a function of the frequency $\omega$ computed via the Petviashvili iteration and numerical continuation for $\eps = 0.20$. The stability of a ground state is investigated numerically by means of an accurate Fourier-type spectral scheme \cite{Boyd2000, Trefethen2000}, see also \cite{Fedele2011}. We found that smooth ground states are stable in agreement with the criterion formulated by \cite{Vakhitov1973}, since $\od{\A}{\omega} > 0$ (see also \cite{Zakharov2001, Yang2010}). Further, we notice that an abrupt reduction in the action $\A$ occurs at a critical frequency $\omega_c(\eps)$ and solitons with wedge-type singularities, viz. peakons, bifurcate from a smooth solitary waves as shown in the right panel of Figure \ref{fig:fig1}. Clearly, this bifurcation can be interpreted as a possible indication of the non-existence of smooth solitons above the critical threshold $\omega_c$. As one can see, as $\omega$ increases the soliton shape tends to become asymmetric and steeper until the critical threshold $\omega_c = 0.85$, above which the smooth solitary waves bifurcate to peakons, which are unstable in agreement with \cite{Vakhitov1973}, since $\od{\A}{\omega} < 0$. Peakons also bifurcate from smooth solitons of given frequency $\omega$ as the steepness $\eps$ increases, as clearly seen in Figure \ref{fig:fig1_5}. Note that in both cases ground states grow asymmetrically before bifurcating to smaller peakons, and so do travelling waves ($c \neq 0$). However, such bifurcation is not observed in the 2-D NLS or Dysthe equations (see, for example, \cite{Fedele2011}), as clearly seen in Figures \ref{fig:fig2_6} -- \ref{fig:fig2_8} where we report the  dependence of  the three invariants $\A$, $\M$ and $\H$ on the parameters $\omega$, $c$ and $\eps$ for the cDZ and cDZ-Dysthe equations respectively.

\begin{figure}
  \centering
  \includegraphics[width=0.99\textwidth]{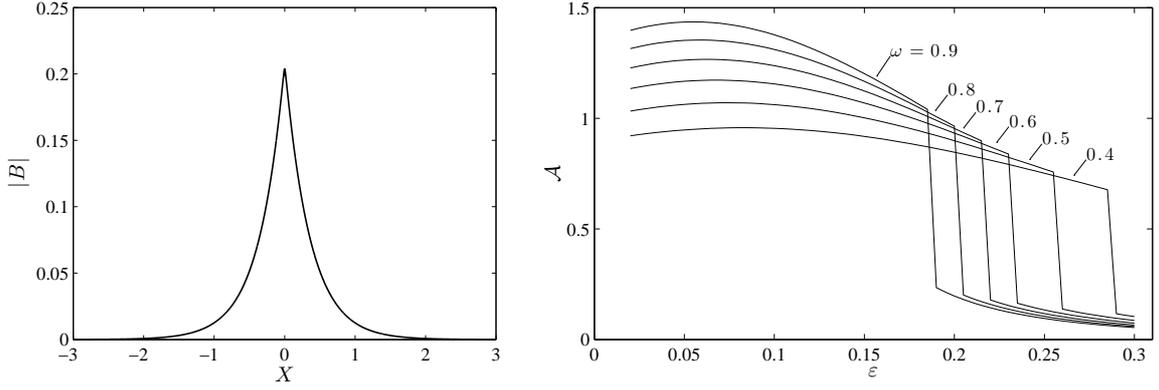}
  \caption{cDZ peakon: (left) numerical solution via the Petviashvili scheme ($\eps = 0.2$, $c = 0$, number of Fourier modes $N \sim 1.5\times 10^6$) and (right) action of ground states of the envelope $B$ as function of the steepness $\eps$ for different values of the frequency $\omega$.}
  \label{fig:fig2}
\end{figure}

\begin{figure}
  \centering
  \includegraphics[width=0.99\textwidth]{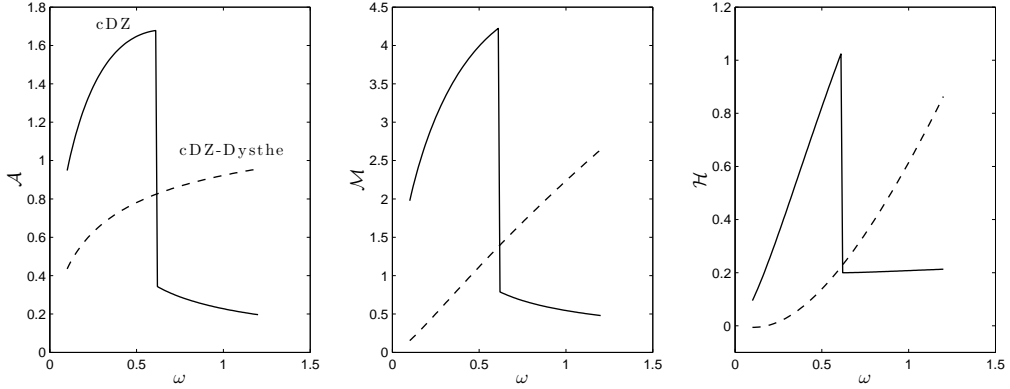}
  \caption{cDZ equation: Action $\A$, momentum $\M$ and Hamiltonian $\H$ dependence on the frequency $\omega$ for ground states of the envelope $B$ ($\eps = 0.23$, $c = 0$).}
  \label{fig:fig2_6}
\end{figure}

\begin{figure}
  \centering
  \includegraphics[width=0.99\textwidth]{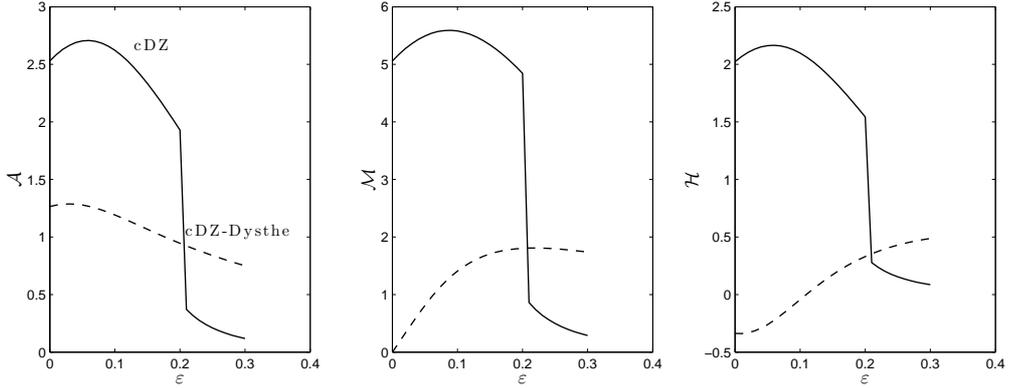}
  \caption{cDZ equation: Action $\A$, momentum $\M$ and Hamiltonian $\H$ dependence on the steepness parameter $\eps$ for ground states of the envelope $B$ ($\omega = 0.80$, $c = 0$).}
  \label{fig:fig2_7}
\end{figure}

\begin{figure}
  \centering
  \includegraphics[width=0.99\textwidth]{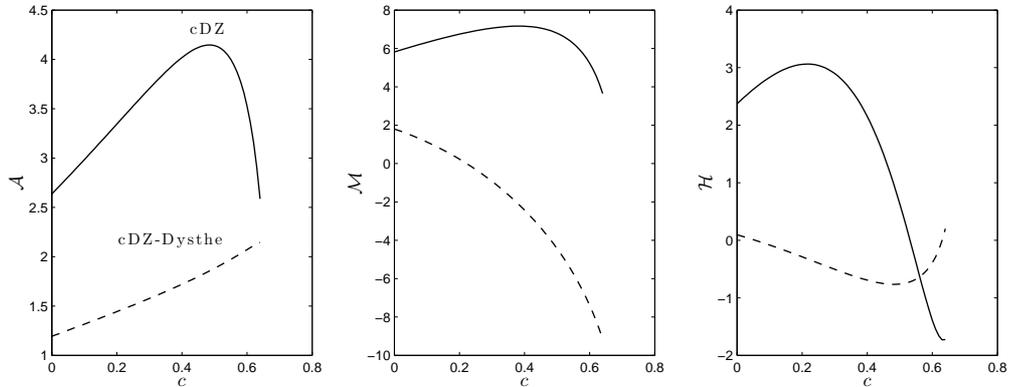}
  \caption{cDZ equation: Action $\A$, momentum $\M$ and Hamiltonian $\H$ dependence on the velocity $c$ for travelling waves of the envelope $B$($\eps = 0.12$, $\omega = 0.90$).}
  \label{fig:fig2_8}
\end{figure}

Note that stable and elastic peakons have been discovered in a special limit of the integrable CH equation \cite{Camassa1993}. On the contrary, from numerical investigations of the cDZ equation peakons appear unstable. The left panel of Figure \ref{fig:fig2} shows the numerically converged peakon obtained via the Petviashvili scheme using $N \sim 1.5 \times 10^{6}$ Fourier modes for $\eps = 0.20$.

\begin{figure}
  \centering
  \includegraphics[width=0.99\textwidth]{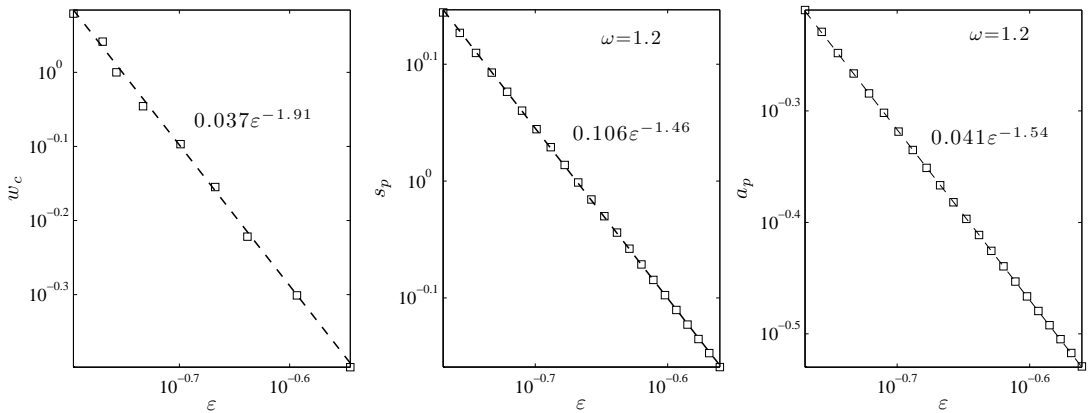}
  \caption{cDZ peakon: dependence of the critical frequency $\omega_c$, peakon slope $s_p$ and amplitude $a_p$ on the steepness parameter $\eps$. This result is obtained for ground states of the envelope $B$. Note that the same scalings hold true also for travelling waves.}
  \label{fig:fig2_9}
\end{figure}

As the frequency $\omega$ increases, peakons bifurcate at smaller steepness $\eps$ as clearly seen in the right panel of Figure \ref{fig:fig2}. Indeed, for ground states we observed numerically that $\omega_c(\eps) \sim \eps^{-2}$ (see left panel of Figure \ref{fig:fig2_9}), and the same scaling holds also for travelling waves ($c \neq 0$). In the physical frame $(x,t)$, owing to the scaling $T = \eps^2\omega_0 t$, bifurcation occurs at a well defined frequency $\omega = \eps^2\omega_c\approx 0.037\omega_0$, which is independent of $\eps$. Further, in the domain $X$ both the peakon amplitude $a_p = \left.|B|\right|_{X=0^-}$ and the associated slope of the wedge singularity (corner)
\[
  s_p = \frac{1}{a_p}\dX\left.|B|\right|_{X=0^-} = \cot(\theta/2)
\]
scales as $\eps^{-3/2}$, $\theta$ being the interior angle of the peakon's singularity (see central and right panels of Figure \ref{fig:fig2_8}). In the physical frame $x$, from (\ref{eq:etaEnv}) the slope $s_p$ becomes 
\[
  s_p = \eps a_0\dX\left.|B|\right|_{X=0^-}\dx X = \eps^3 a_p s_p \approx \cot(170^\circ/2),
\]
which is independent of $\eps$ and corresponds to an angle $\theta \approx 170^\circ$. For a meaningful comparison of this result with the almost-highest solitary wave theory of Longuet-Higgins and Fox \cite{Longuet-Higgins1977, Longuet-Higgins1978, Longuet-Higgins1996}, which predicts an angle of $\theta \approx 120^\circ$, higher order corrections to (\ref{eq:etaEnv}) hidden within the full canonical transformation of \cite{Dyachenko2011} should be accounted for. This is currently under study and will be discussed elsewhere. However, it is also important to mention that the cDZ equation is derived assuming weak nonlinearities, so the strongly nonlinear peakon could be a non-physical artifact, whereas the almost-highest wave satisfies the full Euler equations and thus accounts for nonlinearities of all orders.

\begin{figure}
  \centering
  \subfigure%
  {\includegraphics[width=0.49\textwidth]{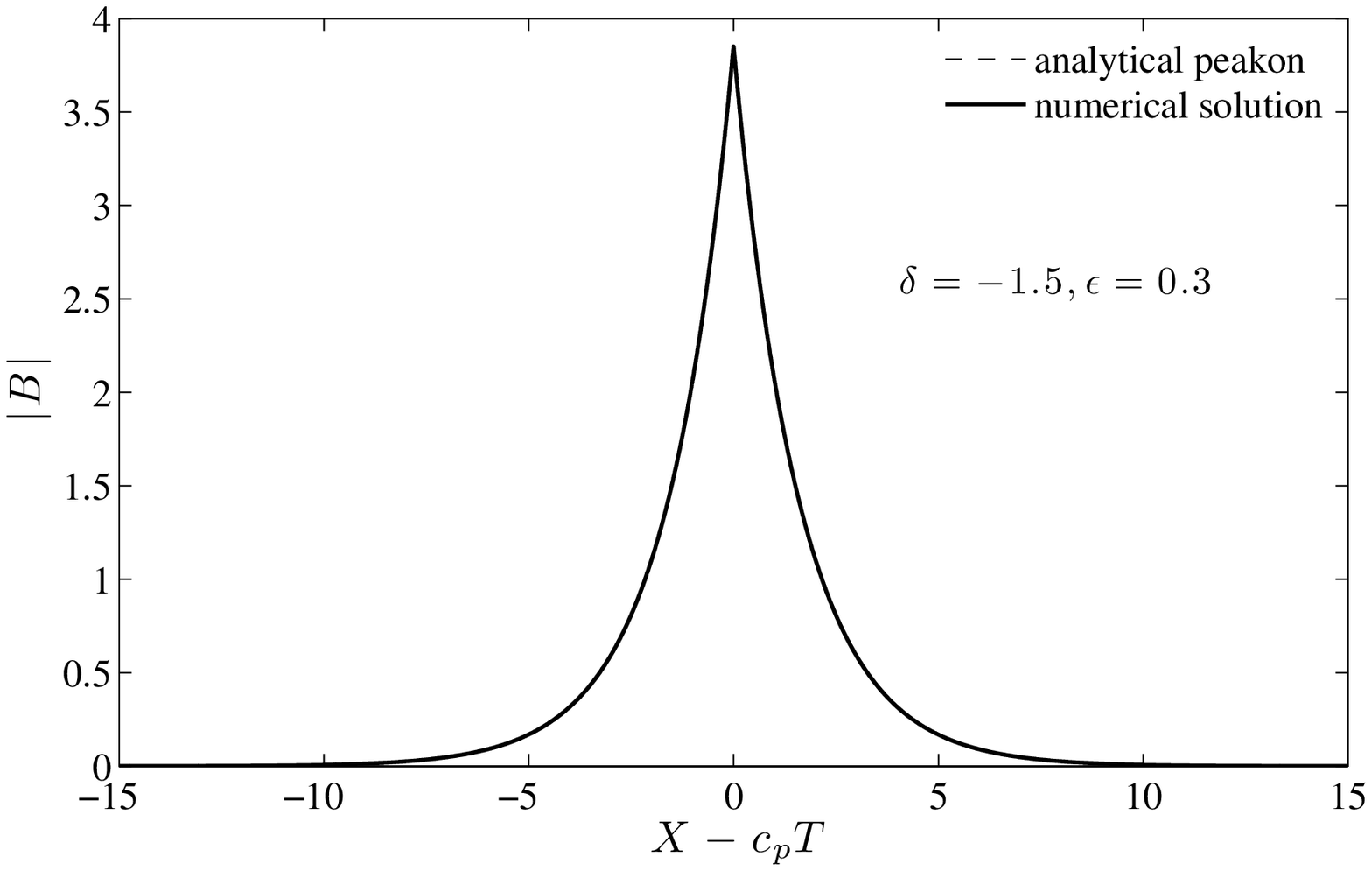}}
  \subfigure%
  {\includegraphics[width=0.49\textwidth]{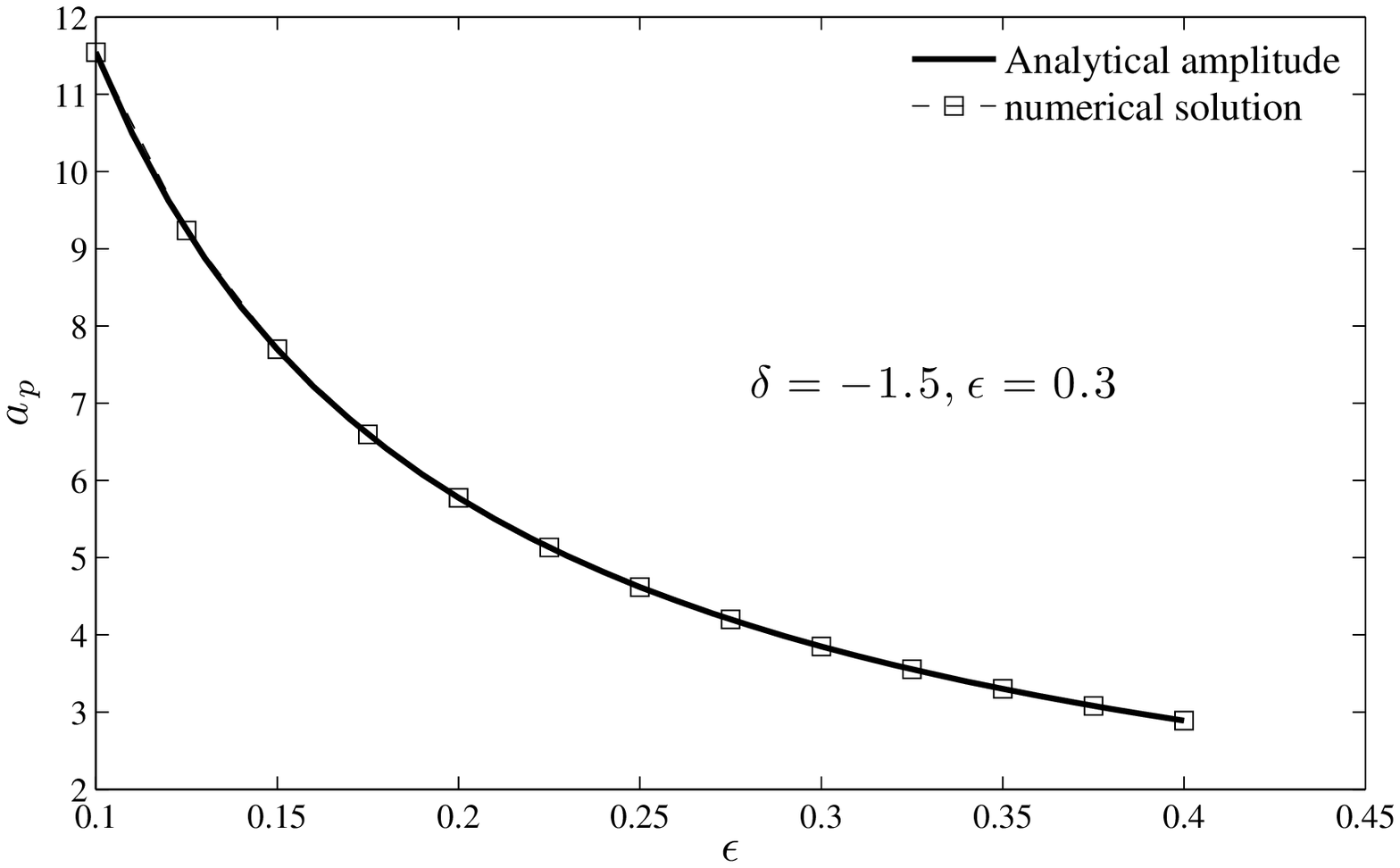}}
  \caption{Left: numerical peakon (solid) of the perturbed local cDZ equation (\ref{eq:lDZ}) and associated analytical solution (\ref{eq:peakon}) (dash) for $\eps = 0.30$, $\delta = -1.5$; Right: amplitude of the numerical peakon against its theoretical counterpart (\ref{eq:amplp}) as function of $\eps$.}
  \label{fig:fig2_99}
\end{figure}

The derivation of an analytical solution for peakons of the non-local cDZ equation (\ref{eq:BT}) is a challenge. To gain more insights into the properties of these singular solutions we consider the academic case of a perturbed version of the local cDZ, viz.
\begin{equation}\label{eq:lDZ}
  iB_T = \frac{1}{8}B_{XX} + \frac{i}{4}\bigl(B^\ast\S((\S B)^2) + (1+\delta)iB^\ast(\S B)^2 - 2\S\bigl(B|\S B|^2\bigr)\bigr),
\end{equation}
Here, the non-local terms are dropped simply because peakons with exponential shape can be found analytically as function of the free parameter $\delta$. Indeed, the ansatz
\begin{equation}\label{eq:peakon}
  B(X,T) = a_p e^{i(\kappa X - \omega_p T)} e^{-\alpha^2|X-c_p T|},
\end{equation}
satisfies (\ref{eq:lDZ}) for 
\begin{equation}\label{eq:peakan}
  \kappa = -\frac{8+\delta}{8\eps}, \qquad
  \alpha^2 = \frac{|\delta|}{8\eps}, \qquad
  \omega_p = -\frac{4+\delta}{32\eps^2}, \qquad
  c_p = \frac{8+\delta}{32\eps}.
\end{equation}
The amplitude $a_p$ is still unknown and it rules the existence of peakons. Indeed, they arise as a balance between the dispersion $\frac{1}{8}B_{XX}$ and the $O(\eps^2)$ term of the nonlinearity $\frac{i}{4}B^\ast\S((\S B)^2)$ in (\ref{eq:lDZ}). These two terms are interpreted in distributional sense because they give rise to Dirac delta functions that must vanish by properly chosing the amplitude $a_p$, thus satisfying the differential equation (\ref{eq:lDZ}) in the sense of distributions. As a result
\begin{equation}\label{eq:amplp}
  a_p = \frac{1}{\eps}\sqrt{-\frac{2}{\delta}}.
\end{equation}
Clearly, peakons exist for $\delta < 0$ and for $\delta = 0$ they reduce to periodic wave solutions of the local cDZ. Thus, peakons can bifurcate from periodic waves of the local cDZ as $\delta$ changes sign, and ground states occur for $\delta = -8$. We note that both the NLS and Dysthe equations do not have higher nonlinear terms similar to $\frac{i}{4}B^\ast\S((\S B)^2)$ and thus they cannot support peakons. We conclude that wedge-type singular solutions are special features of the cDZ equation. The left panel of Figure \ref{fig:fig2_99} shows the remarkable agreement of a numerical peakon ($\delta = -3/2$ and $\eps = 0.3$) computed using the Petviashvili scheme (Fourier modes $N \sim 1.5\times 10^6$) and the corresponding analytical solution (\ref{eq:peakon}). Moreover, the amplitude of the numerical peakon agrees with the associated theoretical counterpart (\ref{eq:amplp}) as function of $\eps$ (see right panel of Figure \ref{fig:fig2_99}). Finally, note that the analytical peakon (\ref{eq:peakon}) has a slope that scales as $\eps^{-1}$, in contrast to the $\eps^{-3/2}$ scaling observed numerically for the full cDZ singular solution. In the physical frame $x$, the interior angle of the wedge singularity  is independent of $\eps$ and so is the peakon frequency $\omega_p$, in agreement with the numerical cDZ peakons. This is an indirect evidence of the validity of our numerical results for the cDZ equation. Finally, we note that as $\eps$ approaches zero, (\ref{eq:lDZ}) tends to the NLS equation, and all the peakon parameters approach infinite values as an indication of the non-existence of singular solutions in the NLS limit.

\begin{figure}
  \centering
  \includegraphics[width=0.89\textwidth]{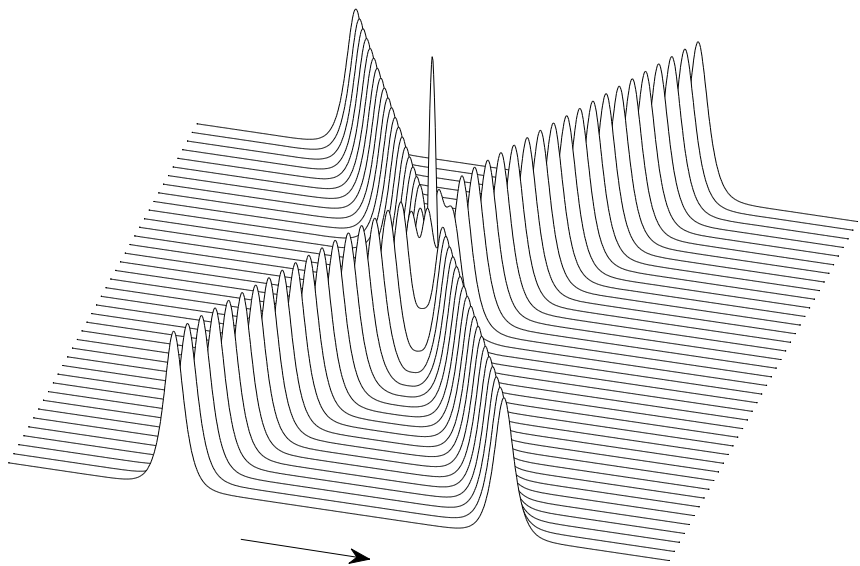}
  \caption{cDZ envelope equation: collision of two smooth travelling waves ($\eps = 0.10$).}
  \label{fig:fig3}
\end{figure}

\begin{figure}
  \centering
  \includegraphics[width=0.99\textwidth]{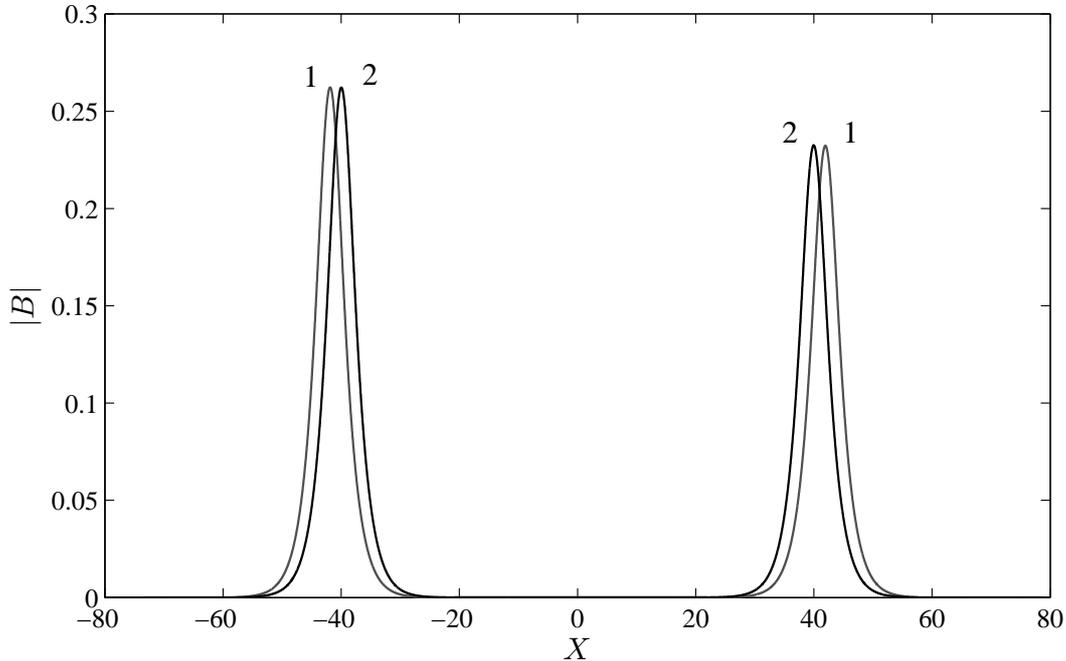}
  \caption{cDZ envelope equation: Initial (1) shape and (2) after collision of the two interacting traveling waves of Figure \ref{fig:fig3}.}
  \label{fig:fig4}
\end{figure}

\begin{figure}
  \centering
  \includegraphics[width=0.89\textwidth]{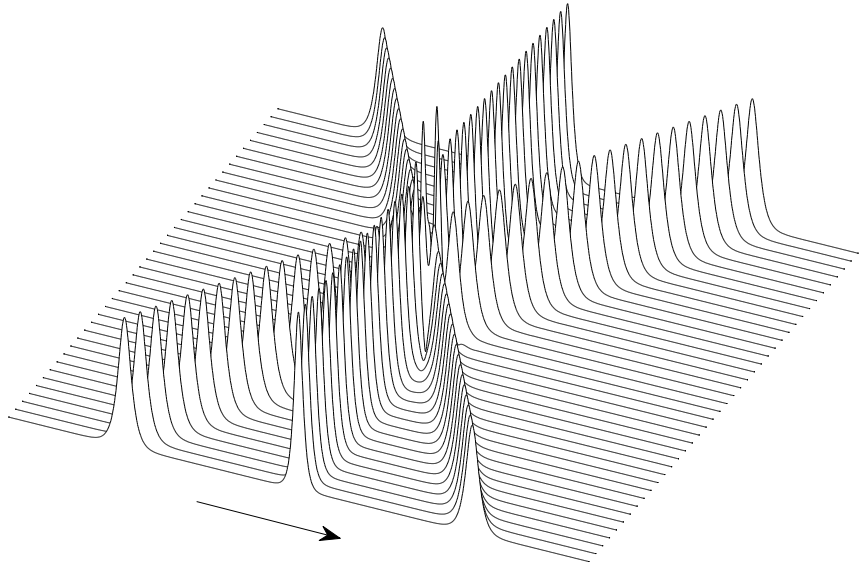}
  \caption{cDZ envelope equation: Elastic collision of two traveling waves with a ground state ($\eps = 0.10$).}
  \label{fig:fig5}
\end{figure}

\begin{figure}
  \centering
  \includegraphics[width=0.99\textwidth]{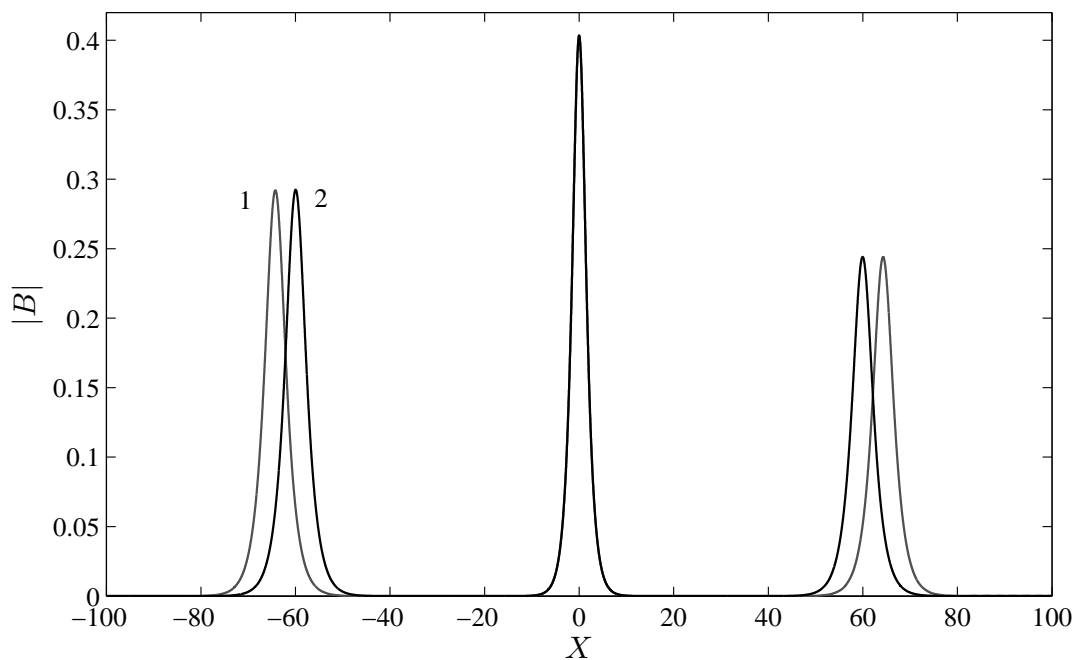}
  \caption{cDZ envelope equation: Initial (1) shape and (2) after collision of the three solitons of Figure \ref{fig:fig5}.}
  \label{fig:fig6}
\end{figure}

\begin{figure}
  \centering
  \includegraphics[width=0.99\textwidth]{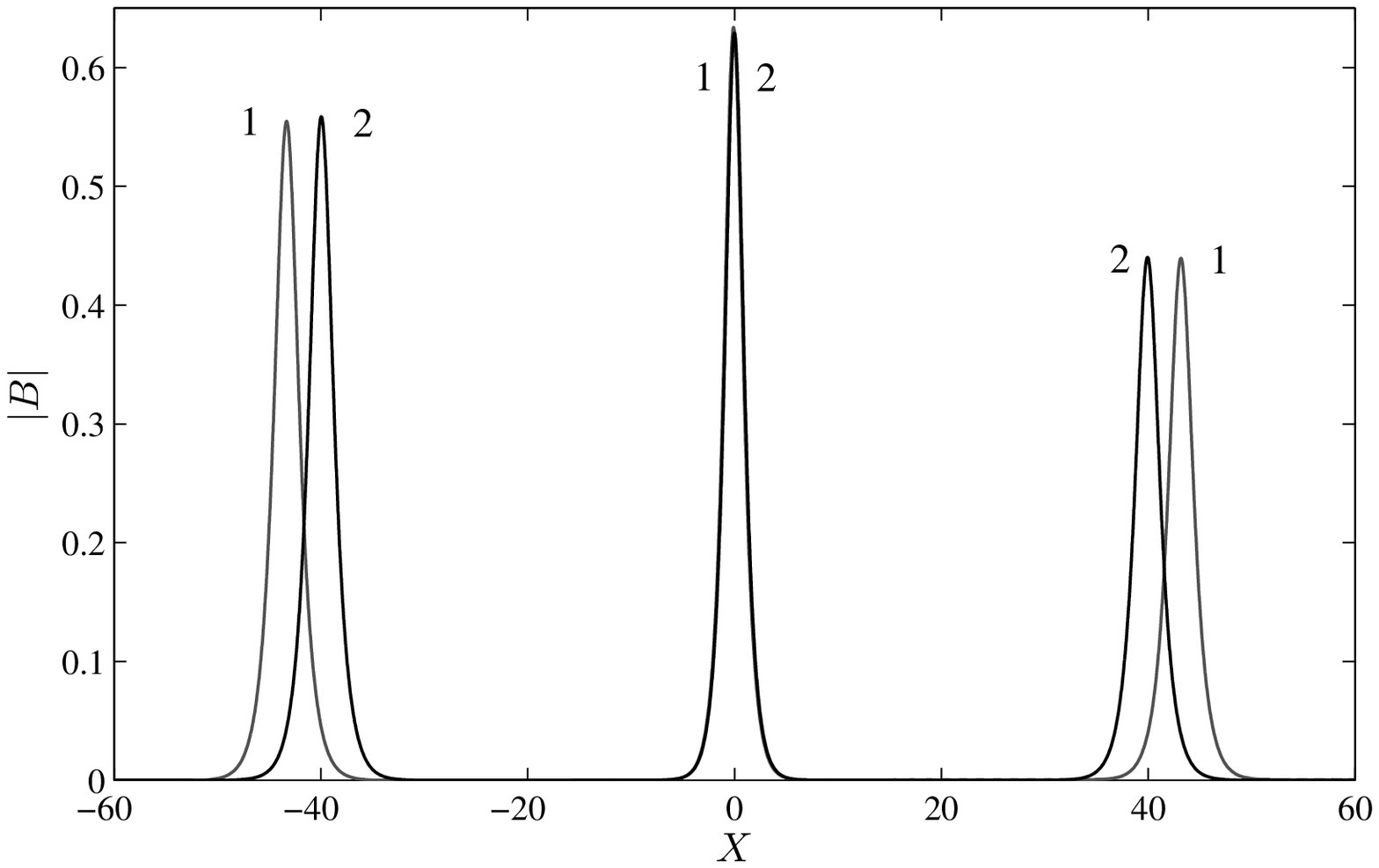}
  \caption{cDZ-Dysthe envelope equation: Initial (1) shape and (2) after collision of two traveling  waves and a ground state ($\eps = 0.10$).}
  \label{fig:fig7}
\end{figure}

\section{Traveling wave interactions}

Hereafter, we investigate the collision of smooth traveling waves of the cDZ equation (\ref{eq:cDZenv}) by means of a highly accurate Fourier-type pseudo-spectral method. We will first briefly describe the adopted spectral approach and then discuss the numerical results.

\subsection{Numerical method description}

We rewrite (\ref{eq:cDZenv}) in the operator form
\begin{equation}\label{eq:Op}
  B_t + \L\cdot B = \N(B),
\end{equation}
where $\L = i\frac{1}{8}\partial_{XX}$ and
\[
  \N(B) = \frac{1}{4}\bigl(B^\ast\S((\S B)^2) + iB^\ast(\S B)^2 - 2\S\bigl(B|\S B|^2\bigr)\bigr) + \frac{\eps i}{2}\Bigl[B\K\{|\S B|^2\} - \S\bigl(\S B\K\{|B|^2\}\bigr)\Bigr].
\]
We solve (\ref{eq:Op}) numerically by applying the Fourier transform in the space variable $X$. The transformed functions will be denoted by $\hat{B} = \F\{B\}$. We recall that the symbol of the non-local term is equal to $|k|$, and that of $\L$ is $i\frac{1}{8}k^2$, $k$ being the Fourier transform parameter. The nonlinear terms are computed in physical space, while space derivatives $\partial_X$ and the nonlocal operator $\K = \Hilb[\partial_x]$ are computed in Fourier space. For example, the term $B^2B_X$ is discretized as
\[
  \F\{B^2 B_X\} = \F\{\bigl(\F^{-1}(\hat{B})\bigr)^2\cdot\F^{-1}\{ik\hat{B}\}\},
\]
and all nonlinear terms are treated in a similar way. We note that the usual 4/3 rule is applied for anti-aliasing since we have to deal with cubic nonlinearities \cite{Trefethen2000, Clamond2001, Fructus2005}.

In order to improve the stability of the space discretization procedure, we can integrate exactly the linear terms. This is achieved by making a change of variables \cite{Milewski1999, Fructus2005}:
\[
  \hat{W}_t = e^{(t-t_0)\L}\cdot\N\Bigl\{e^{-(t-t_0)\L}\cdot\hat{W}\Bigr\},
   \qquad 
  \hat{W}(t) \equiv e^{(t-t_0)\L}\cdot\hat{B}(t), \qquad 
  \hat{W}(t_0) = \hat{B}(t_0).
\]
Finally, the resulting system of ODEs is discretized in space by the Verner's embedded adaptive 9(8) Runge--Kutta scheme \cite{Verner1978}. The step size is chosen adaptively using the so-called H211b digital filter \cite{Soderlind2003, Soderlind2006} to meet the prescribed error tolerance, set as of the order of machine precision.

\subsection{Numerical results}

In all the performed simulations the accuracy has been checked by following the evolution of invariants (\ref{eq:H}), (\ref{eq:AM}). From a numerical point of view the cDZ equation becomes gradually stiffer as the steepness parameter $\eps$ increases, or if higher order dispersion terms are included. Consequently, the conservation of invariants $\H$, $\A$ and $\M$ might be degraded. Nevertheless, even in the stiffest situations a decent accuracy was assured by choosing a sufficiently large number of Fourier modes and the dispersion operator $\Omega_\eps = \frac18\partial_{XX}$. For example, for $\eps = 0.10$ the number $N = 16384$ of Fourier modes were sufficient to achieve conservation of the invariants close to $\sim 10^{-13}$. As an application, consider the interaction between two solitary waves traveling in opposing directions with the same speed ($\eps = 0.10$). The plot of Figure \ref{fig:fig3} shows that the two solitons emerge out of the collision with the same shape, but a slight phase shift. The interaction appears elastic as clearly seen from the plot of Figure \ref{fig:fig4}, which reports the initial and final shapes of the two solitary waves. Further, the interaction of two traveling waves with a ground state appears also elastic (see Figures \ref{fig:fig5} -- \ref{fig:fig6}). This suggests the integrability of the cDZ equation (\ref{eq:cDZenv}) in agreement with the recent results of \cite{Dyachenko2012}. However, the associated Hamiltonian version of the Dysthe equation (\ref{eq:cDZDysthe}) does not support elastic collisions as shown in Figure \ref{fig:fig7}.

\section{Conclusions}

Special solutions of the compact Zakharov equation in the form of traveling waves are numerically constructed using the Petviashvili method. The stability of ground states agrees with the Vakhitov-Kolokolov criterion. Further, unstable ground states with wedge-type singularities, viz. peakons, are numerically identified bifurcating from smooth ground states.  As an academic case, we considered a perturbed version of the local form of the cDZ equation, for which an analytical solution of peakons with exponential shape is derived. Finally, by means of an accurate Fourier-type pseudo-spectral scheme, it is also shown that smooth solitary waves appear to collide elastically, suggesting the integrability of the compact Zakharov equation, but not that of the associated Hamiltonian version of the Dysthe equation.

\section*{Acknowledgements}

D.~\textsc{Dutykh} acknowledges the support from French Agence Nationale de la Recherche, project MathOc\'ean (Grant ANR-08-BLAN-0301-01). F.~\textsc{Fedele} acknowledges the travel support received by the Geophysical Fluid Dynamics (GFD) Program to attend part of the summer school at the Woods Hole Oceanographic Institution in August 2011 and 2012. F.~\textsc{Fedele} thanks Prof. Gregory \textsc{Eyink} for useful discussions on weak solutions of partial differential equations. The authors are also grateful to Profs. Mike \textsc{Banner}, Didier \textsc{Clamond}, Taras \textsc{Lakoba}, Paul \textsc{Milewski}, and Jianke \textsc{Yang} for useful discussions on the subject of nonlinear waves.

\bibliography{biblio}
\bibliographystyle{plain}

\end{document}